\documentclass[%
 reprint,
 amsmath,amssymb,
 aps,
]{revtex4-1}

\usepackage{graphicx}
\usepackage{dcolumn}
\usepackage{bm}

\begin{document}

\preprint{APS/123-QED}

\title{\boldmath Heavy $Q\bar Q$ ``Fireball" Annihilation
       to Multiple Vector Bosons}

\author{Wei-Shu Hou}
 \affiliation{Department of Physics, National Taiwan University, Taipei, Taiwan 10617}


\begin{abstract}
Drawing analogy of replacing
the nucleon by heavy chiral quark $Q$,
the pion by Goldstone boson $G$,
and $\pi NN$ coupling by $GQQ$ coupling,
we construct a statistical model for
$Q\bar Q \to nG$ annihilation,
i.e. into $n$ longitudinal weak bosons.
This analogy is becoming prescient since the LHC direct
bound $m_Q > 611$ GeV implies strong Yukawa coupling.
Taking $m_Q \in (1,\ 2)$ TeV,
the mean number $\langle n_G \rangle$ ranges from 6 to over 10,
with negligible two or three boson production.
With individual $t'$ or $b'$ decays suppressed either by phase space
or quark mixing, and given the strong Yukawa coupling,
$Q\bar Q\to nV_L$ is the likely outcome for
very heavy $Q\bar Q$ production at the LHC.
\begin{description}

\item[PACS numbers]
14.65.Jk 
\end{description}
\end{abstract}

\pacs{Valid PACS appear here}
\maketitle


\section{INTRODUCTION}

Despite the hint for a light Higgs boson at 125 GeV~\cite{125},
there has been keen interest in the search of
new heavy chiral quarks at the Large Hadron Collider (LHC),
resulting in the stringent limit of $m_Q > 611$ GeV/c$^2$~\cite{bpCMS12}.
This is already above the perturbative, tree-level
partial wave unitarity bound (UB) that is
nominally around 550 GeV/c$^2$~\cite{Chanowitz78}.
Thus, if such heavy quarks exist, their Yukawa couplings
would already be in the strong coupling regime.
With TeV scale heavy quark masses,
the actual UB violation (UBV) in the high energy limit
for $Q\bar Q$ scattering may be out of reach.
Instead, the question to ask is:
\emph{Should the current search strategy for
ultraheavy quark $Q$ at the LHC be modified?}
In this note we draw on the analogy of the proton
to argue that $Q\bar Q \to nG$ ($G \equiv V_L$ is
the longitudinal component of the vector boson)
may be the new signature at the LHC.

The $\pi NN$ coupling $g_{\pi NN}^2/4\pi \simeq 14$~\cite{piNN}
gives $g_{\pi NN} \simeq 13$, which is very large and
quite close to the $\pi NN$ ``Yukawa coupling",
$\lambda_{\pi NN} \equiv \sqrt{2}m_N/f_\pi \simeq 14$.
Although $\lambda_Q \equiv \sqrt{2}m_Q/v \gtrsim 3.5$
($v \cong 246$ GeV is the electroweak symmetry breaking scale)
from the current $m_Q$ bound is not yet as large,
drawing analogy with $p\bar p$ annihilation,
we expect that $Q\bar Q \to nG$ may be the dominant process
for $m_Q \in (1,\ 2)$ GeV.

\section{\boldmath Phenomenology of $p\bar p \to n\pi$}

Let us briefly review the observed phenomena regarding
$p\bar p$ annihilation, which is well known~\cite{Klempt05, Dover92}
to go mainly via a ``fireball" into $n$ pions.
The salient features of the annihilation ``fireball" are
(see Fig.~1):
\begin{itemize}
\item Size of order $1/m_\pi$;
\item Temperature $T \simeq 120$ MeV;
\item Average number of emitted pions $\langle n_\pi \rangle \simeq 5$;
\item A soft-pion $p_\pi^2/E_\pi^2$ factor
 modulates the Maxwell--Boltzman distribution for the pions.
\end{itemize}

\begin{figure}[b!]
\centering
\vspace{-1mm}
 {\includegraphics[width=80mm]{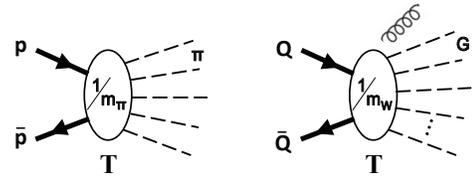}
 }
\vskip-36mm
\caption{
Illustration for $p\bar p \to n\pi$ for $n = \langle n_\pi \rangle \cong 5$,
and the analog of $Q\bar Q \to nG$, where $G = V_L$ is the
Goldstone boson of electroweak symmetry breaking.
Depending on $m_Q = 1$--2 GeV, $\langle n_G \rangle$
could go from 6 to 12 (Eq.~(\ref{PnG})).
The gluon line is to indicate the relatively
soft shedding of color.
} \label{QQ-to-nG}
\end{figure}

It is worthwhile to elucidate these features a little further.
The size $1/m_\pi$ means that the $p\bar p$ annihilation
system, destined to shed the $p$ and $\bar p$ content,
extends over a region $\sim 1/m_\pi$.
The system seems to thermalize to a temperature of
order 120 MeV, hence ``loses memory" of its origins,
and the emitted pions carry momenta that satisfy a
thermal distribution.
This rapid thermalization probably takes place due to the
rather large $\pi NN$ (as well as $\pi\pi$) coupling,
while the $p_\pi^2/E_\pi^2$ suppression~\cite{Orfanidis73-1}
(satisfied rather well by data; see Fig.~12 of Ref.~\cite{Klempt05})
for low pion momentum from the thermal distribution reflects
the Goldstone nature of pion couplings. That is,
the $\pi$ as Goldstone boson couples derivatively,
hence cannot get emitted at zero momentum.
This seems to explain the enhancement factor of
1.3 for the mean kinetic energy $\langle K_\pi \rangle
 \equiv \langle E_\pi \rangle - m_\pi$ beyond
equipartition expectation of $\frac{3}{2}\, T$.
Thus, the relatively high
$\langle E_\pi \rangle \sim 370$ MeV gives rise to
$\langle n_\pi \rangle \simeq 5.1$, as compared to the
maximal allowed number of pions, $2m_N/m_\pi \simeq 13.4$.

\begin{table*}[t!]
\begin{center}
\begin{tabular}{c|cccccccccccccccccc}
  \hline\hline
  $P(n)\;\backslash\; n$ & 2 & 3 & 4 & 5 & 6 & 7 & 8 & 9 & 10 & 11 & 12 & 13 & 14 & 15 & 16 & 17 & 18 \\
  \hline
  $P_{p\bar p}$ & \ 0.4\% \ & \ 8\% \ & \ 18\% \ & \ 46\% \ & \ 22\% \ & \ 6\% \ & \ 0.3\% \ & $-$ & & & & & & & & & & \\
  $P_{Q\bar Q_1}$ & \ 0.1\% \ & \ 1\% \ & \ 6\% \ & \ 19\% \ & \ 31\% \ & 27\% & 12\% & 3\% & 0.4\% & $-$ & & & & & & &  \\
  $P_{Q\bar Q_2}$ & & & & & $-$ & 0.2\% & 0.9\% & \ 3\% \ & \ 8\% \ & \ 16\% \ & \ 22\% \ & \ 22\% \
                  & \ 16\% \ & \ 8\% \ & \ 3\% \
                                        & \ 0.9\% \ & \ 0.2\% \\
  \hline\hline
\end{tabular}
 \caption{Sample multiplicity distributions:
     $P_{p\bar p}(n)$ is the observed distribution for $p\bar p \to n\pi$~\cite{Klempt05},
     while $P_{Q\bar Q_{1\,(2)}}(n)$ is the
     $Q\bar Q \to nG$ distribution for $m_Q = 1\ (2)$ TeV
     according to Eq.~(\ref{PnG}), where
     $G \equiv V_L$ is the electroweak Goldstone boson.}
\end{center}
\end{table*}

At a more refined level, it is found that
$\langle n_{\pi^\pm} \rangle \simeq 3.1$ and
$\langle n_{\pi^0} \rangle \simeq 2.1$, with
$2\langle n_{\pi^0} \rangle/\langle n_{\pi^\pm} \rangle > 1$,
i.e. more neutral pions are emitted than charged ones.
Furthermore, the pion multiplicity distribution appears Gaussian,
\begin{equation}
 P(n_\pi) = \frac{1}{\sqrt{2\pi}\,\sigma}\,
            e^{-{(n_\pi - \langle n_\pi \rangle)^2 / 2\sigma^2}},
 \label{Pnpi}
\end{equation}
with $\sigma \sim 1$. More specifically~\cite{Orfanidis73-2},
\begin{equation}
 \sigma \simeq \frac{1}{2} \sqrt{\langle n_\pi \rangle},
 \label{sigma}
\end{equation}
is argued from statistical models~\cite{Jabs71}.
Thus, $\sigma \simeq 1.13$ gives a good fit to
data~\cite{Klempt05}, which is given in Table~I.
Note the rather small $p\bar p \to \pi\pi$ 2-body fraction, and
the cutoff of pion multiplicity above 8.

This successful ``statistical model" which accounts for
gross features of $p\bar p \to n\pi$ annihilation
goes back to Fermi~\cite{Fermi50}, who considered
a system of noninteracting pions.
It has been refined through the years, and
the strong interactions of the pions do play a role.
One final aspect is a focusing of incoming waves
by attractive potential that leads to strong absorption
in a smaller region than originally suggested.

\section{\boldmath $Q\bar Q \to nV_L$ Analog}

We mean by $Q$ a left-handed chiral doublet
(with corresponding right-handed weak singlets) that is
degenerate in mass, thereby possessing a
\emph{heavy isospin} symmetry $I_Q$,
much like the nucleon $N$.
This is nothing but the 4th generation~\cite{4S4G}.
To draw true analogy with the $\pi NN$ case,
the $GQQ$ Yukawa coupling $\lambda_Q$ should be
of order 13--14, i.e. $m_Q \gtrsim 2$  TeV. However,
we will assume that analogous phenomena already
appears for 1 TeV, hence we will consider
$m_Q \in (1,\ 2)$ TeV.

With Higgs mechanism already established,
the Goldstone boson $G \equiv V_L$
carries a length $1/M_W$~\cite{IQ},
which defines the \emph{size} of the
$Q\bar Q$ annihilation fireball.
Besides the dynamical mechanism for
$m_p$ and $m_Q$ generation, comparing
$m_\pi \propto m_u + m_d$ with $M_W \propto g$,
where $g$ is the weak gauge coupling,
the size of the fireball is in part
determined by unrelated, ``random" parameters.

The fireball temperature $T \equiv T_{Q\bar Q \to nG}$
is harder to assess. Noting that
$T_{p\bar p \to n\pi} \sim 120\ {\rm MeV}
 < T_c^{\rm QCD} \sim 170\ {\rm MeV}$,
likely $T < T_c^{\rm EW}$, where $T_c^{\rm EW}$ is
the electroweak transition temperature.
By this analogy, however, one notes that
$T_c^{\rm QCD}$ arises from the detailed
underlying theory for hadron phenomena
(which includes $p\bar p \to n\pi$), QCD.
Even though we believe EW symmetry breaking (EWSB)
probably~\cite{Hou12} arises from strong Yukawa coupling,
$\lambda_Q$, we do not yet have an
underlying theory for $\lambda_Q$ itself.
Thus, we do not have a good handle on $T$, except that
it is in the 100 GeV scale, of order $v$.
We shall therefore take as nominal
\begin{equation}
 T \sim \frac{2}{3}\, v \sim 160\ {\rm GeV},
 \label{T}
\end{equation}
which can be interpreted as either $1.3 \times \frac{1}{2}\, v$
(here 1.3 corresponds to $T_{p\bar p \to n\pi}/f_\pi$),
or $v\times T_{p\bar p \to n\pi}/T_c^{\rm QCD}$.
The latter would give 170 GeV, which is
not so different from Eq.~(\ref{T}).
We stress, however, that the fireball temperature
could be 1.5, even twice as high, and should be
determined eventually by experiment.
The Goldstone $p_G^2/E_G^2$ factor
should still modulate the thermal $p_G$ distribution.
But because of the smallness of $M_W^2$ compared with $4m_Q^2$,
the modulation is considerably milder than the $p\bar p \to n\pi$ case,
so $\langle K_G \rangle$ should be closer to $\frac{3}{2}\, T$.

Assuming Eq.~(\ref{T}) but without applying the
1.3 enhancement factor over equipartition (as is
the case for $p\bar p \to n\pi$), we take
$\langle K_G \rangle \sim \frac{3}{2}\, T \sim 240$ GeV,
hence $\langle E_G \rangle \sim 320$ GeV, or
\begin{equation}
 \langle |p_G| \rangle \sim 310\ {\rm GeV},
 \label{pG}
\end{equation}
with $\gamma_G \sim 4$.
For $m_Q = 1\ (2)$ TeV, or $2m_Q = 2\ (4)$ TeV,
this corresponds to
\begin{equation}
 \langle n_G \rangle \sim 6.25\ (12.5),
 \label{nG}
\end{equation}
where we artificially keep three digits of significance
for generating a ``realistic" multiplicity distribution.
Assuming Eqs.~(\ref{Pnpi}) and (\ref{sigma}), we have
$\sigma \simeq \sqrt{\langle n_G \rangle}/2 \sim 1.25\ (1.77)$,
and the multiplicity distribution is
\begin{equation}
 P(n_G) \simeq 0.319\, e^{-\frac{(n_G - 6.25)^2}{3.13}}\ \;
         \left(0.226\, e^{-\frac{(n_G - 12.5)^2}{6.25}}\right),
 \label{PnG}
\end{equation}
for $m_Q = 1\ (2)$ TeV.
We note that a higher fireball temperature $T$ would
result in lower $\langle n_G \rangle$,
higher $\langle |p_G| \rangle$ and
a narrower distribution (controlled by $\sigma$).

We illustrate the $Q\bar Q \to nG$ process in Fig.~1
(gluon emission discussed later),
and tabulate the multiplicity distributions in Table I.
For $m_Q = 1$ TeV, about 90\% of $Q\bar Q$
annihilations go into 5--8 prongs of $V_L \equiv G$.
Several $V_L$s should be considerably above 300 GeV momentum, while
4-prong events (at 6\%) are in general composed of
$V_L$s with momentum $\sim$ 500 GeV.
Therefore, $W$-tagged ``fat" jets, $j_W$, should become
a useful tool for identifying these multi-$V_L$ events.
For $m_Q = 2$ TeV, again over 90\% of $Q\bar Q$ annihilations
go into 10--15 prong $V_L$s, which is a rather large number.
For 9--12 prong events (at $\sim 50\%$),
a significant number of $V_L$s would have
momentum above 400 GeV, while for higher multiplicity,
many should still carry momentum higher than the mean,
Eq.~(\ref{pG}).
These high multiplicity $nV_L$ events would be
possible hallmark for heavy $Q\bar Q$ production.

\section{Production and Competing Modes}

If our analogy with $p\bar p$ annihilation is
already realized for $m_Q = 1$ TeV, then
even at 8 TeV running of LHC, where of order
15 fb$^{-1}$ data is expected in 2012,
one could already get a hint.
The cross section is of order a couple fb,
so one might observe some number of 4 or more
$W$-tagged jet ($j_W$) events, with
additional jet multiplicity that are less well
$W$-tagged. The competing modes would be
regular $Q\bar Q$ production, followed by
``free quark decay", e.g. (assuming $m_{b'} > m_{t'}$)
$b'\bar b' \to t\bar tW^+W^- \to b\bar bWWWW$,
or $t'\bar t' \to b\bar bW^+W^-$~\cite{AH06};
we shall assume CKM hierarchy for simplicity.
We see that the $W$-jet multiplicity is lower,
associated with isolated high $p_T$ $b$-jets,
and practically no $Z$-jets~\cite{jZ1} or $Z \to \ell^+\ell^-$.
In contrast, $Q\bar Q \to nV_L$ does not have
isolated $b$-jets ($b$-jets would come in pairs
at lower fraction, to form a $j_Z$ from $Z\to b\bar b$),
$W$-jet multiplicity is higher, and tend to
have $Z$-jets~\cite{jZ2}.
We expect the $Q\bar Q \to nV_L$ fireball process
would dominate over the
$Q\bar Q \to b\bar bWW(WW)$ free quark decay process,
as we would argue shortly.

There are arguments that, if the heavy chiral quarks
$Q$ themselves are responsible~\cite{Hou12} for EWSB,
then $m_Q > 1$ TeV is likely~\cite{MHK}.
Our earlier analogy with the $\pi NN$ ``Yukawa" coupling
suggests $m_Q \sim 2$ TeV. If so, the prospect
for the 2012 LHC run at 8 TeV is not good, and
one would have to wait for the 13--14 TeV run,
expected by late 2014.
Running the HATHOR code~\cite{Hathor} for $Q\bar Q$ production,
we estimate the 14 TeV cross sections to be
of order 50--60 fb for $m_Q = 1$ TeV,
dropping to $\sim 3$ fb for $m_Q = 1.5$ TeV,
and 0.2--0.3 fb for $m_Q = 2$ TeV.
From $2m_Q = 2$ to 4 TeV,
one quickly runs out of parton luminosity.
Note that $q\bar q \to Q\bar Q$ production dominates over
$gg \to Q\bar Q$ production, as the valence quark supplies
the needed large parton momentum fraction.

From the cross section and expected LHC luminosities,
for $m_Q \lesssim 1.5$ TeV,
again we do not foresee a problem for discovery.
Note that, assuming $I_Q$ symmetry,
i.e. near degeneracy of $t'$ and $b'$,
then
\begin{itemize}
\item $t'\; (b') \to b'\; (t') + W^*$ decay:

Suppressed by both phase space and small Goldstone momentum;
\item $t'\; (b') \to b\; (t) + W$:

Suppressed by CKM element $|V_{t'b}|$ ($|V_{tb'}|$).

With no sign of New Physics in $B_s \to J/\psi\phi$,
$B_s \to \mu^+\mu^-$, and $B_d \to K^{*0}\mu^+\mu^-$,
one expects~\cite{HKX} such CKM elements to be less than 0.1.
\end{itemize}
In contrast, once $q\bar q \to Q\bar Q$ pulls the
heavy quark pair out of the vacuum,
the $Q\bar Q$ pair ``sees" a cross section of order $1/M_W^2$,
which is at the $\mu$b level.
With $q\bar q \to Q\bar Q$ production,
there is Yukawa attraction~\cite{EHY} between $Q\bar Q$
that mimics the focusing attraction for $p\bar p \to n\pi$.
Thus, there is good likelihood that
$Q\bar Q \to nG$, i.e. $nV_L$, would dominate
over free quark decay.

We comment that the produced $Q\bar Q$ is likely
in a color-octet state, hence in general
it would need to shed color.
However, gluons have no way to sense the $T \sim 160$ GeV
(or higher) of the fireball, which is of electroweak nature.
Instead, the heaviness of $Q$ means gluon radiation
is $1/m_Q$ suppressed (heavy quark symmetry).
We illustrate gluon radiation in Fig.~1,
but expect the associated gluon-jet to be soft
and does not provide a discriminant.
Since the fireball is viewed as a nonperturbative
process, it is hard to assess
how this gluon is actually radiated.

\section{Discussion and Conclusion}

A natural question is whether the annihilation 
gets modified by heavy $Q$ motion.
Here, $p\bar p$ annihilation data again provide a guide.
Fig.~1 of Ref.~\cite{Dover92} illustrates, for example,
that for $p_{\bar p} < m_p/4$ (lab frame),
the annihilation cross section $\sigma_{\rm ANN}$
predominates the total cross section $\sigma_{\rm TOT}$.
In fact, $\sigma_{\rm ANN}$ always dominate over
the elastic cross section $\sigma_{\rm EL}$ even for 
$p_{\bar p}$ greater than several times $m_p$.
Only for $p_{\bar p} \gtrsim 3m_p$ does the 
inelastic $\sigma_{\rm PROD}$ become significant,
dominating beyond $5m_p$ or so.
For our purpose, we expect annihilation, 
in the way we discussed, to be dominant
before the motion turns rather relativistic.
If one really has a collider of much higher energy,
then $Q\bar Q$ scattering becomes an issue related
to UBV. If the analog to $\sigma_{\rm EL}^{\bar pp}$ 
implies subsequent free $Q$ decay, 
then one might still observe free quark decay. 
The question of how $Q$ decays in large Yukawa coupling 
limit needs to be investigated nonperturbatively.

In case $m_Q \gtrsim 1.5$ TeV,
one quickly runs out of parton luminosities
(higher energy would be preferred!), hence one would
need high luminosity running of LHC at 14 TeV.
However, the situation need not be so pessimistic:
the very large Yukawa coupling suggests the
existence of bound states below $2m_Q$.
For example, as discussed in Ref.~\cite{EHY}, there is
likely an isosinglet, color-octet $\omega_8$ resonance
that can be produced via $q\bar q \to \omega_8$.
How $\omega_8$ decays would depend on more details
of the $Q\bar Q$ bound state spectrum and properties.
The beauty of our analogy with $p\bar p \to n\pi$ annihilation
is precisely the thermal nature of this
fireball process~\cite{Klempt05,Dover92},
with little ``remembrance",
either of the initial $p\bar p$ state,
or detailed resonances in the hadron spectrum.
Thus, we make no assertion on
$\omega_8$ decay properties here, except that it
offers hope for an enhanced production cross section.

If the decay of the $\omega_8$ is analogous to
the fireball picture, then by $m_{\omega_8} < 2m_Q$
and the resonance production nature,
there is good hope for earlier discovery.
If $\omega_8$ decays through similar chains
as discussed in Ref.~\cite{EHY}, then it might lead to
the discovery of several resonances.
The study of Ref.~\cite{EHY} was done with
500 GeV $< m_Q < 700$ GeV in mind, to avoid
issues of boundstate collapse~\cite{Hou12}.
But since this region is now close to being ruled out,
a numerical update, in particular also on
obtaining the spectrum, is certainly called for.
This would require nonperturbative solutions
for strong Yukawa coupling.

An offshoot study of Ref.~\cite{EHY} provides an
interesting contrast.
If free quark decay is suppressed by very small $V_{t'b}$,
and some kinematic selections are operative,
it is argued that $\omega_8 \to \pi_8 + W$
($\pi_8$ is some isotriplet, color-octet Yukawa-bound ``meson"),
followed by $\pi_8 \to W_T + g$, where $W_T$ is transverse,
with the upshot of $\omega_8 \to W W g$.
This is an exception to our fireball discussion,
in that
 1)~it is effectively 2-body in vector bosons ($WW$);
 2)~the gluon is energetic.
If reconstructed~\cite{AEHY}, one could discover
{\it two} resonances.
These signatures arise from special conditions
that are unlikely to hold in general.

Multiple weak boson production has been considered
below $Q\bar Q$ threshold~\cite{HM90}.
There is a rise of high multiplicities
as one starts to approach threshold of high $m_Q$.
But that would be out of the range of validity
for the $gg \to nG$ amplitude via virtual $Q$ loop
considered.
We remark that our multi-$V_L$ signature is in principle
quite distinct from micro-blackhole production~\cite{bhCMS12}.
Micro-blackholes in essence emit all types
of particles democratically. In contrast, our
fireball is heated in the electroweak sense,
and by far prefers emitting the strongly coupled
weak Goldstone bosons $V_L \equiv G$.
However, since searches so far are based only on the
simplified signature of high jet multiplicities,
a refined search is needed to separate micro-blackholes
from $Q\bar Q$ fireballs.

A final remark is in regards $V_LV_L$ scattering.
If a very heavy chiral quark doublet $Q$
exists above the TeV scale, it would not be easily
compatible with a light Higgs because of the very
large quadratic corrections to the light Higgs mass.
A corollary of our argument would then suggest that
the traditional $V_LV_L \to V_LV_L$ scattering
study for heavy Higgs case may be the wrong
place to search for New Physics enhancement.
Instead, one should again watch out for $V_LV_L$
scattering to high(er) multiplicity of $V_L$s~\cite{AAD12}.
In general, one should treat the UBV in $V_LV_L$
and $Q\bar Q$ scattering as one single problem.

In conclusion,
the ever-increasing mass bound on heavy
sequential chiral quark $Q$,
and the associated question of whether there might be
a need to change the search strategy, prompt us to draw analogy
with the observed $p\bar p \to n\pi$ ``fireball" annihilation.
We suggest that, in the range of $m_Q \in (1,\ 2)$ TeV,
$Q\bar Q$ might annihilate via an analogous fireball
into $n$ Goldstone, or longitudinal vector bosons,
with mean multiplicity $\langle n_G \rangle$
ranging from $\sim 6$ to over 10.
The mean multiplicity $\langle n_G \rangle$ is correlated
with the fireball ``temperature" $T$,
which are the key parameters for experiment to measure.
This process would at least dilute the
usual free quark decay picture for
on-shell $Q\bar Q$ production.
Strong Yukawa boundstates below $2m_Q$ should further aid
the discovery of new heavy chiral quarks, and
the prospect appears optimistic beyond the unitarity bound.

\vskip0.3cm
\noindent{\bf Acknowledgement}.
We thank Johan Alwall, Kai-Feng Chen and Eberhard Klempt for discussions,
and Ta-Wei Wang for help on HATHOR.
This research is supported by NSC 100-2745-M-002-002-ASP
and various NTU grants under the MOE Excellence program.


\begin{thebibliography}{9}
\bibitem{IQ}
  Assuming $I_Q$ symmetry, we take $M_Z = M_W$ and ignore weak
  gauge couplings $g$, $g'$ compared with $\lambda_Q$.
\bibitem{jZ1}
  It is not our purpose to explore the separation
  of $j_Z$ from $j_W$. Certainly this would be
  difficult just by ``jet mass". We note, however,
  that $j_W$ should not carry a $b$-tag. In contrast,
  over 1/4 of $Z$ decays are into $b\bar b$ (15\%) or
  $c\bar c$ (12\%). Given that $W\to c\bar s$ is about 1/3,
  the $c\bar c$ component should probably be rejected.
  Thus, if one could double $b$-tag a fat-jet with
  high efficiency and purity, $j_Z$ might be selectable.
  This would enrich the rare $Z\to e^+e^-$ and $Z\to \mu^+\mu^-$.
  On a seperate note, if all $j_W$ are hadronically selected,
  in principle a recoiling non-observed ``missing-$Z$"
  (considering the 20\% fraction of $Z\to \nu\bar\nu$)
  could perhaps be pursued.
\bibitem{jZ2}
  The analogy with $p\bar p \to n\pi$ suggests a slight
  excess of $\pi^0$ over (half the) $\pi^\pm$.
  However, the $Z$ mass is heavier than the $W$,
  so this point is to be determined by experiment.
\end{thebibliography}
\end{document}